\documentclass[12pt]{article}
\usepackage{a4wide,epsfig,amsmath,amssymb,cite}
\usepackage{scalefnt}
\newcommand{\Lmu}{L_\mu}

\newcommand{\order}[1]{\mathcal{O}\left( #1 \right)}

\begin{document}


\title{\vskip-3cm{\baselineskip14pt
\begin{flushleft}
\normalsize DESY 06-107 \\
\normalsize SFB/CPP-06-31 \\
\normalsize TTP06-22 \\
\end{flushleft}}
\vskip1.5cm
Fermionic Corrections to the Three-Loop Matching Coefficient of the
Vector Current
}

\author{\small 
  P. Marquard$^{(a)}$, J.H. Piclum$^{(a,b)}$, D. Seidel$^{(a)}$,
  M. Steinhauser$^{(a)}$
\\
{\small\it (a) Institut f{\"u}r Theoretische Teilchenphysik,
  Universit{\"a}t Karlsruhe (TH)}\\
{\small\it 76128 Karlsruhe, Germany}
\\
{\small\it (b) II. Institut f\"ur Theoretische Physik, 
  Universit\"at Hamburg}\\
{\small\it 22761 Hamburg, Germany}
}

\date{}

\maketitle

\thispagestyle{empty}

\begin{abstract}
In this paper we consider the matching coefficient of the vector
current between Quantum Chromodynamics (QCD) and Non-Relativistic QCD
(NRQCD) to three-loop order in perturbation theory.
We evaluate the fermionic corrections containing a closed massless
fermion loop.
The results are building blocks both for the bottom and top quark system
at threshold. We explain in detail the methods used for the evaluation
of the Feynman diagrams, classify the occurring master integrals and
provide results for the latter. The numerical effects are significant. 
They have the tendency to improve the behaviour of the perturbative
series --- both for the bottom and top quark system.

\medskip

\noindent
PACS numbers: 12.38.Bx 14.65.Ha
\end{abstract}

\newpage


\section{Introduction}

One of the main goals of a future international linear collider (ILC) is
the precise measurement of the top quark threshold. 
Next to a precise extraction of the strong coupling an unrivaled
determination of the top quark mass and its width is possible.
This would open up a new chapter in the electroweak precision physics
which leads to very strong checks of the Standard Model or
possible extensions.

The theoretical calculations are based on an effective
theory~\cite{Caswell:1985ui,Bodwin:1994jh} (for a review
see~\cite{Brambilla:2004wf})
which enables the simultaneous expansion in the two small
parameters present in threshold phenomena, the strong coupling
$\alpha_s$ and the velocity of the heavy quarks, $v$. 
Furthermore, a resummation of correction terms like $(\alpha_s/v)^n$
to all orders is possible in perturbation theory.
In a first step
the effective theory is constructed in a matching procedure where it
is required that the Green functions agree with the corresponding ones
in full QCD in the limit of infinitely heavy top quark mass.
In a second step the Green function in the effective theory has to be
evaluated leading to the desired cross section.

Experimental studies~\cite{Martinez:2002st} have shown that an
uncertainty of less 
than 3\% is necessary for the theoretical predictions for the cross
section of the reaction $e^+ e^-\to t\bar{t}$ in the threshold region,
i.e. for $\sqrt{s}\approx 350$~GeV.
This has not yet been achieved.

As far as the strong coupling is concerned
currently the analysis is complete to next-to-next-to-leading
order (NNLO)~\cite{Hoang:2000yr} demonstrating the urgent need for the 
third-order corrections.
This huge enterprise has been started already some years ago and many 
building blocks have been evaluated starting from the determination of
the third-order Hamiltonian~\cite{Kniehl:2002br}, third-order
corrections to the energy levels and wave
functions~\cite{Penin:2002zv,Penin:2005eu,Beneke:2005hg} up to the
evaluation of the contributions from the Coulomb potential to the
non-relativistic Green function~\cite{Beneke:2005hg}.
Next to fixed-order corrections also the resummation of
next-to-next-to leading logarithmic terms has been considered and quite
some progress has been
achieved~\cite{Pineda:2001ra,Pineda:2001et,Hoang:2003ns,Penin:2004ay}.

In order to reach a theory-uncertainty below 3\% it is also necessary to
worry about electroweak effects which can be parametrically 
as big as the third-order QCD terms. First results in this context
are available~\cite{Hoang:2004tg,Eiras:2006xm} which show that indeed numerical
effects in the region of a few percent are possible.

In this paper a further building block needed for the N$^3$LO
analysis is provided: the fermionic corrections to the matching
coefficient of the vector current. More precisely we compute the
three-loop contributions to the matching coefficient containing at
least one closed loop of massless fermions which we enumerate by $n_l$.

The paper is organized as follows: in the next section we describe in
detail the theoretical framework and the methods used for the
calculation. In particular, we classify the occurring integrals and
discuss their evaluation. Sections~\ref{sec::z2} and~\ref{sec::match} 
contain a detailed discussion about the 
wave function and matching coefficient up to order 
$\alpha_s^3 n_l$, respectively. 
In Section~\ref{sec::phen} we briefly discuss the
phenomenological impact of our result and Section~\ref{sec::concl} contains
our conclusions and outlook.
In the Appendix a complete list of master integrals is provided ---
both for the ``natural'' and the $\epsilon$-finite basis.


\section{Method}

The matching coefficient establishing the relation between the quark
current in the full and effective theory
constitutes a building block for all threshold phenomena involving the
coupling of a photon or $Z$ boson to heavy quarks.
We define the currents through
\begin{eqnarray}
  j_v^\mu &=& \bar{Q} \gamma^\mu Q\,,
  \nonumber\\
  \tilde{j}^i &=& \phi^\dagger \sigma^i \chi\,,
\end{eqnarray}
and generically denote the heavy quark by $Q$.
$\phi$ and $\chi$ are two-component Pauli spinors for quark and
anti-quark, respectively.
The definition of the matching coefficient is established through
\begin{eqnarray}
  j^k_v &=& c_v(\mu) \tilde{j}^k 
  +\frac{d_v(\mu)}{6m_Q^2}\phi^\dagger\sigma^k \vec{D}{}^2\chi
  + \ldots
  \label{eq::def_of_cv}
  \,,
\end{eqnarray}
where $k=1,2,3$ denotes the spacial components. 
$\vec{D}$ contains the space-like
components of the gauge-covariant derivative involving the gluon fields and
the ellipsis stand for operators of higher mass dimension. 
The second term on the r.h.s. of Eq.~(\ref{eq::def_of_cv}) is already of
NNLO with $d_v=1+{\cal O}(\alpha_s)$.
The Wilson coefficients $c_v$ and $d_v$ may be expressed as series
in $\alpha_s$ and represent the contributions from the hard modes
which have been integrated out in order to arrive at the effective theory.
Note that both the higher order terms in the inverse heavy quark
mass and the time-like component is not of interest for the aim of this paper.
The purpose of this paper is the evaluation of the fermionic
correction to the non-singlet contribution of $c_v$ containing at least
one closed massless quark loop.
The one-loop corrections to $d_v$ contain no contribution proportional
to $n_l$.

In order to compute $c_v$ it is obvious to consider the 
$Q\bar{Q}\gamma$ vertex which in the following is denoted by $\Gamma_v$.
This effectively transforms Eq.~(\ref{eq::def_of_cv}) into
\begin{eqnarray}
  Z_2 \Gamma_v &=& c_v \tilde{Z}_2 \tilde{Z}_v^{-1} \tilde{\Gamma}_v
  + \ldots
  \,,
  \label{eq::def_cv}
\end{eqnarray}
where we have on the left- and right-hand side quantities of the full and
effective theory, respectively,
and the ellipses denote terms suppressed by inverse powers
of the heavy quark mass. 
In Eq.~(\ref{eq::def_cv}) it is understood that 
the couplings and masses in $\Gamma_v$ are renormalized.

Since $c_v$ takes into account the degrees of freedom which have been
integrated out from the full theory it depends in our case only 
on $\alpha_s$ and --- for dimensional reasons --- on $\ln\mu^2/m_Q^2$.
In particular, $c_v$ does not explicitly depend on the momenta of the external
particles. Thus, it is useful to apply the so-called threshold
expansion~\cite{Beneke:1997zp,Smirnov:2002pj} to Eq.~(\ref{eq::def_cv})
which has the consequence that $\Gamma_v$ has to be evaluated for 
$s=4m_Q^2$ since all except the hard region cancel in Eq.~(\ref{eq::def_cv}).
Furthermore, on the right-hand side only tree contributions have to be
considered. In particular we have $\tilde{Z}_2=1$.

Starting from two-loop order the matching procedure
exhibits infra-red divergences which are compensated by ultra-violet
divergences of the effective theory rendering physical quantities
finite. In Eq.~(\ref{eq::def_cv}) the renormalization constant
$\tilde{Z}_v$ which generates the anomalous dimension of $\tilde{j}_v$
takes over this part. 
Note that the vector current in the full theory does not get
renormalized. 

\begin{figure}
  \leavevmode
  \epsfxsize=\textwidth
 \epsffile{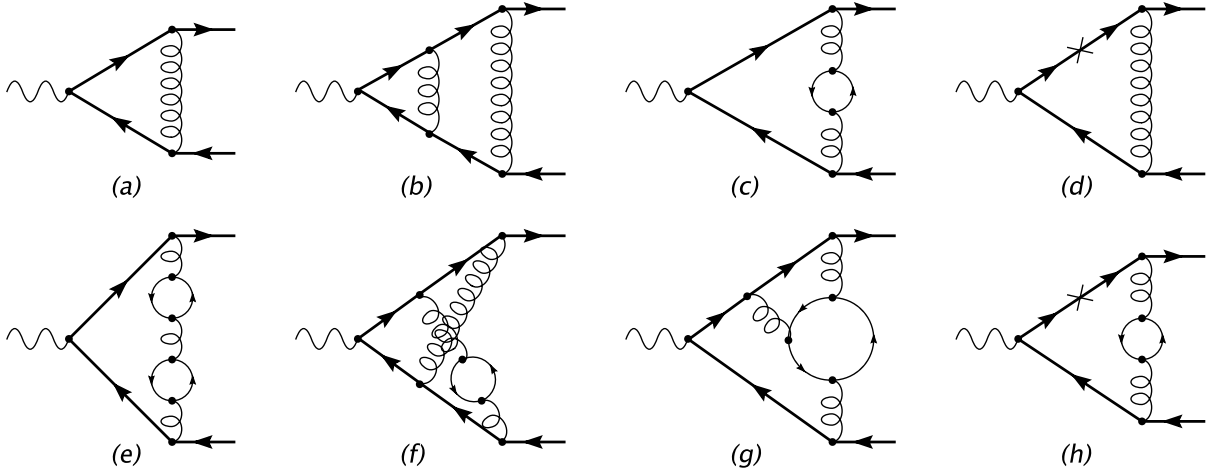}
\caption{\label{fig::sample} Feynman diagrams contributing to the 
  matching coefficient. Bold lines denote heavy quarks with mass $m_Q$,
  thin lines denote massless quarks and curly lines denote gluons. In
  (d) and (h) mass counterterm diagrams are shown.}
\end{figure}

Sample Feynman diagrams contributing to the one-, two- and three-loop 
fermionic part of $c_v$
are shown in Fig.~\ref{fig::sample}. Due to the special kinematic
situation with $s=q^2=4 m_Q^2$ and on-shell heavy quark lines 
it is possible to perform a partial fractioning in the integrands
corresponding to the various types of diagrams and map all occurring
integrals to one of the following functions:
\begin{eqnarray}
  J^{(2)}_\pm(n_1,\dots,n_5) &=& \left(
  \frac{\mu^{2\epsilon}}{i\pi^{d/2}} \right)^2 \int \frac{{\rm d}^d k\;
  {\rm d}^d l}{ (k^2)^{n_1} (l^2)^{n_2} ((k-l)^2)^{n_3} (k^2 + q\cdot
  k)^{n_4} (l^2 \pm q\cdot l)^{n_5}}\nonumber
  \,,
  \\
  L^{(2)}_\pm(n_1,\dots,n_5) &=& \nonumber \\
  \lefteqn{\!\!\!\!\!\! \left(
  \frac{\mu^{2\epsilon}}{i\pi^{d/2}} \right)^2 \int \frac{{\rm d}^d k\;
  {\rm d}^d l}{ (k^2)^{n_1} (l^2)^{n_2} ((k+l)^2 + q\cdot (k+l))^{n_3}
  (k^2 + q\cdot k)^{n_4} (l^2 \pm q\cdot l)^{n_5}} \,,} &&\nonumber
  \\
  J^{(3)}_\pm(n_1,\dots,n_9) &=& \left(
  \frac{\mu^{2\epsilon}}{i\pi^{d/2}} \right)^3 \int \frac{{\rm d}^d k\;
  {\rm d}^d l\; {\rm d}^d p}{ (k^2)^{n_1} (l^2)^{n_2} (p^2)^{n_3}
  ((k-l)^2)^{n_4} ((l-p)^2)^{n_5} ((p-k)^2)^{n_6}}\nonumber \\
  && \times \frac{(l^2 + q\cdot l)^{-n_8}}{ (k^2 + q\cdot k)^{n_7} (p^2
  \pm q\cdot p)^{n_9}}\nonumber
  \,,
  \\
  L^{(3,n_l)}_\pm(n_1,\dots,n_9) &=& \left(
  \frac{\mu^{2\epsilon}}{i\pi^{d/2}} \right)^3 \int \frac{{\rm d}^d k\;
  {\rm d}^d l\; {\rm d}^d p}{ (k^2)^{n_1} (l^2)^{n_2} ((k+l)^2 + q\cdot
  (k+l))^{n_3} (k^2 + q\cdot k)^{n_4}} \nonumber \\
  && \times \frac{(p^2 + q\cdot p)^{-n_9}}{ (l^2 \pm q\cdot l)^{n_5}
  (p^2)^{n_6} ((p+k)^2)^{n_7} ((p-l)^2)^{n_8}}
  \,.
  \label{eq::integrals}
\end{eqnarray}
For convenience we have also listed the two-loop functions
$J^{(2)}_\pm$ and $L^{(2)}_\pm$ originally defined in
Ref.~\cite{Beneke:1997zp}. The three-loop functions 
$J^{(3)}_\pm$ and $L^{(3,n_l)}_\pm$
contain irreducible
scalar products which are shown as numerators in the three-loop
integrals in Eq.~(\ref{eq::integrals}). The corresponding indices can
only adopt values less or equal to zero. Furthermore, only two out of
the three indices $n_6$, $n_7$ and $n_8$ in $L^{(3,n_l)}_\pm$ can have
positive values.
Note that the integrals $J_+^{(2)}$, $L_+^{(2)}$ and $J_+^{(3)}$ are
actually two-point functions whereas the integrals $J_-^{(2)}$,
$L_-^{(2)}$, $J_-^{(3)}$ and $L^{(3,n_l)}_\pm$ correspond to vertices.

All Feynman diagrams are generated with {\tt
QGRAF}~\cite{Nogueira:1991ex}. The various topologies are identified 
with the help of {\tt q2e} and {\tt
exp}~\cite{Harlander:1997zb,Seidensticker:1999bb} 
which also adapts the notation in order to match the one of
Eq.~(\ref{eq::integrals}).

In a next step the reduction of the various functions to so-called master
integrals has to be achieved. For this step we have chosen two
approaches: the Laporta method~\cite{Laporta:1996mq,Laporta:2001dd} 
and Baikov's method~\cite{Baikov:1996rk,Baikov:1996iu} in
the formulation of~\cite{Smirnov:2003kc}. The application of these
methods reduces the three-loop integrals to twelve master
integrals. Since some of the master integrals are only known numerically
it is very useful to construct a so-called $\epsilon$-finite
basis~\cite{Chetyrkin:2006dh}. 
Details of this procedure and the results for the master integrals are
given in Appendix~\ref{masters}.
Let us in the following provide more details on our implementation of 
each of the two reduction methods.


\subsection{Implementation of the Laporta algorithm}

The Laporta algorithm is based on the
integration-by-parts (IBP) relations~\cite{Chetyrkin:1981qh} where in a first step
a huge system of equations is created by inserting numerical integer values 
into the relations. After assigning a weight specifying the complexity
to each occurring integral the system of equations is solved
step-by-step. In this way an arbitrary integral is expressed in terms
of master integrals which cannot be further reduced.

To date there is only one publicly available computer code, 
{\tt AIR}~\cite{Anastasiou:2004vj}, where the Laporta algorithm is
implemented. We have successfully applied {\tt AIR} to the two-loop
diagrams. However, the three-loop integrals of
Eq.~(\ref{eq::integrals}) cannot be treated with the help of {\tt AIR}. 

For this reason we have chosen to write a new implementation of Laporta's
algorithm, {\tt Crusher}~\cite{PMDS}. It is written in {\tt C++} and uses
{\tt GiNaC}~\cite{Bauer:2000cp} for simple
manipulations like taking derivatives of polynomial quantities. In the
practical implementation of the Laporta algorithm one of the most
time-consuming operations is the simplification of the coefficients
appearing in front of the individual integrals. This task is performed
with the help of {\tt Fermat}~\cite{fermat} where a special interface
has been used (see Ref.~\cite{Tentyukov:2006ys}).
The main features of the
implementation are the automated generation of the IBP identities
and a complete symmetrization of the diagrams. Using {\tt Crusher} we
solved approximately forty million equations to reduce all integrals to master
integrals. 



\subsection{Baikov's method}
Baikov's method is also based on the IBP
relations. Here, however, one explicitly constructs the coefficient
functions of the master integrals as parametric integrals over some
polynomial, which encodes the topology of the initial integral. The
integrations can then be performed as Cauchy integrations around the
origin or as contour integrations between the roots of the
polynomial. The results are rational functions depending on the
dimension and the kinematical invariants.

In some cases, however, it is not possible to perform the integrations
in this way. In these cases one has to solve the
integration-by-parts relations for the parametric integrals. They are
similar to the relations for the initial integral, however, they depend on
less indices and are thus in general significantly simpler. 
Since the relations can still be quite complicated, it is
convenient to again use Laporta's algorithm. In the current calculation
the program {\tt AIR} was used for this task.

In order to perform the calculations, the algorithms for each integral
were implemented in {\tt FORM}~\cite{Vermaseren:2000nd}. In a
first step, however, everything was implemented in {\tt
  Mathematica}. This is very convenient since it is possible
to produce almost all needed {\tt Mathematica} code automatically. Once
the program is finished, it can be easily translated to {\tt FORM} and
the {\tt Mathematica} version can be used for debugging.

The two-loop calculation was done entirely with this method. In the
three-loop case, however, it turned out that a complete solution of the recurrence
relations for the coefficient functions is not possible with {\tt
  AIR}. The calculation was therefore done with the Laporta method
described above. We used Baikov's method only for the $n_l^2$ part and
to cross-check some of the coefficient functions.



\section{\label{sec::z2}Wave function renormalization constant}

According to Eq.~(\ref{eq::def_cv}) the wave function renormalization
constant in the on-shell scheme $Z_2$ constitutes a crucial input for
the computation of $c_v$. It has been computed to two-
and three-loop approximation in Refs.~\cite{Broadhurst:1991fy} 
and~\cite{Melnikov:2000zc}, respectively.
We have repeated the calculation of the two-loop and fermionic
three-loop contributions which are needed for the present calculation
and find complete agreement with the literature. For completeness we
repeat the result for $Z_2$ which can be cast into the form
\begin{eqnarray}
  Z_2 &=& 1 + \frac{\alpha_s(\mu)}{\pi} \left(\frac{e^{\gamma_E}}{4 \pi}
  \right)^{-\epsilon} \delta Z_2^{(1)} +
  \left(\frac{\alpha_s(\mu)}{\pi}\right)^2 \left(\frac{e^{\gamma_E}}{4 \pi}
  \right)^{-2\epsilon} \delta Z_2^{(2)} \nonumber \\
  && + \left(\frac{\alpha_s(\mu)}{\pi}\right)^3 \left(\frac{e^{\gamma_E}}{4
  \pi} \right)^{-3\epsilon} \left( \delta Z_2^{(3,nl)} + \mbox{non-$n_l$
  terms} \right) + \mathcal{O}\left(\alpha_s^4\right) \,,
  \label{eq::Z2}
\end{eqnarray}
with
\begin{eqnarray}
  \delta Z_2^{(1)} &=& - C_F \left[ \frac{3}{4\epsilon} + 1 +
  \frac{3}{4} \Lmu + \left( 2 + \frac{1}{16} \pi^2  + \Lmu + \frac{3}{8}
  \Lmu^2 \right) \epsilon \right. \nonumber \\
  && \left. + \left( 4 + \frac{1}{12} \pi^2 - \frac{1}{4} \zeta(3) +
  \left( 2 + \frac{1}{16}\pi^2 \right) \Lmu + \frac{1}{2} \Lmu^2 +
  \frac{1}{8} \Lmu^3 \right) \epsilon^2 \right] \,,
  \label{eq::Z21l}
\end{eqnarray}
\begin{eqnarray}
  \delta Z_2^{(2)} &=& \left[ \frac{11}{32
  \epsilon^2} - \frac{127}{192 \epsilon} - \frac{1705}{384} +
  \frac{5}{16} \pi^2 - \frac{1}{2} \pi^2 \ln 2 + \frac{3}{4} \zeta(3)
  - \frac{215}{96} \Lmu - \frac{11}{32} \Lmu^2 \right. \nonumber \\
  && + \left( -\frac{9907}{768} + \frac{769}{1152} \pi^2 - \frac{23}{8}
  \pi^2 \ln 2 + \pi^2 \ln^2 2 + \frac{129}{16} \zeta(3) - \frac{7}{40}
  \pi^4 + \frac{1}{2} \ln^4 2 + 12 a_4 \right. \nonumber \\
  && \left.\left. + \left( -\frac{2057}{192} +
  \frac{109}{192} \pi^2 - \pi^2 \ln 2 + \frac{3}{2} \zeta(3) \right) \Lmu -
  \frac{259}{96} \Lmu^2 - \frac{11}{32} \Lmu^3\right) \epsilon \right]
  C_AC_F \nonumber \\
  && + \left[ \frac{9}{32 \epsilon^2} + \left( \frac{51}{64} +
  \frac{9}{16} \Lmu \right) \frac{1}{\epsilon} +
  \frac{433}{128} - \frac{49}{64} \pi^2 + \pi^2 \ln 2 -
  \frac{3}{2} \zeta(3) + \frac{51}{32} \Lmu + \frac{9}{16} \Lmu^2
  \right. \nonumber \\
  && + \left( \frac{211}{256} - \frac{339}{128} \pi^2 +
  \frac{23}{4} \pi^2 \ln 2 - 2 \pi^2 \ln^2 2 - \frac{297}{16} \zeta(3) +
  \frac{7}{20} \pi^4 - \ln^4 2 - 24 a_4 \right. \nonumber \\
  && \left.\left. + \left( \frac{433}{64} - \frac{49}{32} \pi^2 + 2
  \pi^2 \ln 2 - 3 \zeta(3) \right) \Lmu + \frac{51}{32} \Lmu^2 +
  \frac{3}{8} \Lmu^3 \right) \epsilon \right] C_F^2 \nonumber \\
  && + \left[ \left( \frac{1}{16} + \frac{1}{4} \Lmu \right)
  \frac{1}{\epsilon} + \frac{947}{288} - \frac{5}{16} \pi^2
  +\frac{11}{24} \Lmu + \frac{3}{8} \Lmu^2 + \left( \frac{17971}{1728} -
  \frac{445}{288} \pi^2 \right.\right. \nonumber \\ 
  && \left.\left. + 2
  \pi^2 \ln 2 - \frac{85}{12} \zeta(3) + \left( \frac{1043}{144} -
  \frac{29}{48} \pi^2 \right) \Lmu + \frac{5}{8} \Lmu^2 + \frac{7}{24}
  \Lmu^3 \right) \epsilon \right] C_FT \nonumber \\
  && + \left[ -\frac{1}{8 \epsilon^2} + \frac{11}{48 \epsilon} +
  \frac{113}{96} + \frac{1}{12} \pi^2 + \frac{19}{24} \Lmu + \frac{1}{8}
  \Lmu^2 + \right. \nonumber \\
  && \left. + \left( \frac{851}{192} + \frac{127}{288} \pi^2 +
  \zeta(3) + \left( \frac{145}{48} + \frac{3}{16} \pi^2 \right) \Lmu +
  \frac{23}{24} \Lmu^2 + \frac{1}{8} \Lmu^3 \right) \epsilon
  \right] C_FTn_l \,
  \label{eq::Z22l}
\end{eqnarray}
and
\begin{eqnarray}
  \delta Z_2^{(3,nl)} &=&  C_FTn_l \left\{ \left[ \frac{11}{72
  \epsilon^3} - \frac{169}{432 \epsilon^2} + \left( \frac{313}{1296} +
  \frac{1}{4} \zeta(3) \right) \frac{1}{\epsilon}
  \right.\right. \nonumber \\
  && + \frac{111791}{15552} + \frac{13}{48} \pi^2 + \frac{47}{36}
  \pi^2 \ln 2 - \frac{2}{9} \pi^2 \ln^2 2 - \frac{35}{72} \zeta(3) +
  \frac{19}{1080} \pi^4  - \frac{1}{9} \ln^4 2 - \frac{8}{3} a_4
  \nonumber \\
  && \left. +
  \left( \frac{169}{27} - \frac{1}{18} \pi^2 + \frac{1}{3} \pi^2 \ln 2 +
  \frac{1}{4} \zeta(3) \right) \Lmu + \frac{469}{288} \Lmu^2 +
  \frac{11}{72} \Lmu^3 \right] C_A \nonumber \\
  && + \left[ \frac{3}{32 \epsilon^3} + \left( -\frac{19}{192} +
  \frac{3}{32} \Lmu \right) \frac{1}{\epsilon^2} +
  \left( -\frac{235}{384} - \frac{7}{128} \pi^2 - \frac{1}{4} \zeta(3) -
  \frac{41}{64} \Lmu - \frac{3}{64} \Lmu^2 \right) \frac{1}{\epsilon}
  \right. \nonumber \\
  && -\frac{3083}{2304} + \frac{2845}{2304} \pi^2 -
  \frac{47}{18} \pi^2 \ln 2 + \frac{4}{9} \pi^2 \ln^2 2 +
  \frac{473}{96} \zeta(3) - \frac{229}{2160} \pi^4 + \frac{2}{9}
  \ln^4 2 + \frac{16}{3} a_4 \nonumber \\
  &&\left. + \left( -\frac{1475}{384} + \frac{133}{384} \pi^2 -
  \frac{2}{3} \pi^2 \ln 2 + \frac{1}{4} \zeta(3) \right) \Lmu -
  \frac{179}{128} \Lmu^2 - \frac{11}{64} \Lmu^3 \right] C_F \nonumber \\
  && + \left[ \left( \frac{1}{36} + \frac{1}{12} \Lmu \right)
  \frac{1}{\epsilon^2} + \left( -\frac{5}{216} + \frac{1}{144} \pi^2 -
  \frac{1}{9} \Lmu + \frac{1}{24} \Lmu^2 \right) \frac{1}{\epsilon}
  \right. \nonumber \\
  && \left. - \frac{4721}{1296} + \frac{19}{54} \pi^2 - \frac{1}{36}
  \zeta(3) + \left( -\frac{329}{108} + \frac{25}{144} \pi^2 \right)
  \Lmu - \frac{7}{12} \Lmu^2 - \frac{5}{72} \Lmu^3 \right] T \nonumber \\
  && + \left[ -\frac{1}{36 \epsilon^3} + \frac{11}{216 \epsilon^2} +
  \frac{5}{1296 \epsilon} - \frac{5767}{7776} - \frac{19}{108} \pi^2 -
  \frac{7}{18} \zeta(3) \right. \nonumber \\
  && \left.\left. - \left( \frac{167}{216} + \frac{1}{18} \pi^2 \right)
  \Lmu - \frac{19}{72} \Lmu^2 - \frac{1}{36} \Lmu^3 \right] Tn_l \right\}
  \label{eq::Z23lnl}
  \,,
\end{eqnarray}
where $\Lmu = \ln(\mu^2/m_Q^2)$,
$C_F=(N_c^2-1)/(2N_c)$, $C_A=N_c$, $T=1/2$ and $n_l=n_f-1$ is the number of
light-quark flavours. $N_c=3$ and $\alpha_s$ is the strong coupling renormalized
in the $\overline{\rm MS}$ scheme with $n_f$ active flavours.
$\zeta(3)$ is Riemann's $\zeta$ function with the
value $\zeta(3)=1.202057\ldots$, $\gamma_E$ is Euler's constant with the
value $\gamma_E = 0.577216\ldots$ and $a_4 = {\rm Li}_4(1/2) =
0.517479\ldots$.
Note that $\delta Z_2^{(1)}$ and $\delta Z_2^{(2)}$ are given to order
$\epsilon^2$ and $\epsilon$, respectively, which is needed for the
three-loop calculation of $c_v$.

The integrals needed for $Z_2$ constitute a subset of the ones needed
for $c_v$. In fact, from Eq.~(\ref{eq::integrals}) only the integral types
with a ``$+$'' sign contribute. 
We also want to mention that our calculation poses the first
independent check of the $n_l$ part of Ref.~\cite{Melnikov:2000zc}.


\section{\label{sec::match}Matching coefficient}

It is convenient to cast the perturbative expansion of the
matching coefficient in the form
\begin{eqnarray}
  c_v &=& 1 + \frac{\alpha_s(\mu)}{\pi} c_v^{(1)}
  + \left(\frac{\alpha_s(\mu)}{\pi}\right)^2 c_v^{(2)}
  + \left(\frac{\alpha_s(\mu)}{\pi}\right)^3 c_v^{(3)}
  + {\cal O}(\alpha_s^4)
  \,,
  \label{eq::cvdef}
\end{eqnarray}
where the one-~\cite{KalSar} and
two-loop~\cite{Czarnecki:1997vz,Beneke:1997jm} (see also
Ref.~\cite{Kniehl:2006qw}) terms are given by
\begin{eqnarray}
  c_v^{(1)}&=&-2C_F
  \nonumber\\
  c_v^{(2)}&=&\left(-\frac{151}{72}
  +\frac{89}{144}\pi^2
  -\frac{5}{6}\pi^2\ln2-\frac{13}{4}\zeta(3)\right)C_AC_F
  \nonumber\\&&\mbox{}
  +\left(\frac{23}{8}-\frac{79}{36}\pi^2
  +\pi^2\ln2-\frac{1}{2}\zeta(3)\right)C_F^2
  +\left(\frac{22}{9}-\frac{2}{9}\pi^2\right)C_FT
  \nonumber\\
  &&+\frac{11}{18}C_FTn_l
    -\frac{1}{2}\left[4\beta_0+\pi^2\left(\frac{1}{2}C_A +
  \frac{1}{3}C_F\right)\right]C_F\ln\frac{\mu^2}{m_Q^2}
    \,,
  \label{eq::cv12}
\end{eqnarray}
with $\beta_0=(11 C_A/3-4Tn_f/3)/4$. Note that the terms
proportional to $\beta_0 \ln\frac{\mu^2}{m_Q^2}$ are connected to the 
choice $\alpha_s(\mu)$ in Eq.~(\ref{eq::cvdef}) whereas the ones
proportional to $\pi^2 \ln\frac{\mu^2}{m_Q^2}$ originate from the
separation of hard and soft scales in the construction of NRQCD.

The decomposition of $c_v^{(3)}$ according to the colour structures 
is given by
\begin{eqnarray}
  c_v^{(3)} &=& C_F T n_l\left(
  C_F c_{FFL} + C_A c_{FAL} + T c_{FHL} + T n_l c_{FLL} 
  \right)
  + \mbox{non-$n_l$ terms}
  \label{eq::cv3lnl}
  \,.
\end{eqnarray}

In the notation of Eq.~(\ref{eq::def_cv}) it is understood that
in $\Gamma_v$ the renormalization of the coupling constant and heavy
quark mass is already performed. The renormalization of the
coupling constant can be done straightforwardly by replacing the bare
coupling with the renormalization constant times the renormalized
coupling. The mass renormalization, however, is more intricate since we
are dealing with on-shell integrals. It is therefore convenient to
calculate the counterterms directly by considering one- and two-loop
diagrams with zero-momentum insertions in the massive fermion lines. Sample
diagrams are shown in Fig.~\ref{fig::sample}(d) and (h). The vertex
denoted by a cross is replaced by the mass renormalization
constant in the on-shell scheme. One-loop diagrams with two insertions
are not needed at the order considered in this paper.

Starting from three-loop level the wave function renormalization
constant begins to depend on the QCD gauge parameter,
$\xi$~\cite{Melnikov:2000zc}.
However, it is interesting to note that the colour structures entering
our result (cf. Eq.~(\ref{eq::Z23lnl})) are independent of $\xi$. As a
consequence the quantity $\Gamma_v$ also has to be 
independent of $\xi$ which is indeed the case in our calculation.
Actually, the $\xi$ dependence remaining in the sum of all genuine
three-loop diagrams contributing to $\Gamma_v$ is canceled by the
contribution where a two-loop 
mass counterterm is inserted in the one-loop vertex diagram
(cf. Fig.~\ref{fig::sample}(d)).
Note that the mass counterterm contribution from the two-loop
fermionic diagram (cf. Fig.~\ref{fig::sample}(h)) is $\xi$ independent.
The cancellation of the gauge parameter serves as a welcome check for
the correctness of our result.

As already mentioned above, after taking into account all counterterm
contributions there are still infra-red divergences left in the
quantity $Z_2\Gamma_v$ and thus in $c_v\tilde{Z}_v^{-1}$. 
This is a consequence of the threshold expansion
accompanied by dimensional regularization which is used to extract the 
matching coefficient.
Alternatively it would have been
possible to choose a cut-off for the momentum integrations which
finally results in a factorization scale separating the hard and soft momenta.
In our approach the infra-red poles are absorbed into
$\tilde{Z}_v$ which we define in the $\overline{\rm MS}$ scheme.
As a consequence the matching coefficient itself is finite but scale
dependent. 
In the physical quantities this scale dependence gets canceled 
against the corresponding contributions from the effective theory.
Of course, one has to make sure that for the subtraction of the poles
in the effective theory the same scheme is used as in the full theory.

For the renormalization constant $\tilde{Z}_v$ we obtain
\begin{eqnarray}
  \tilde{Z}_v &=& 1 + \left(\frac{\alpha_s(\mu)}{\pi}\right)^2 \left(
  \frac{1}{12} C_F^2 + \frac{1}{8} C_FC_A \right) \frac{\pi^2}{\epsilon}
  \nonumber \\
  && + \left(\frac{\alpha_s(\mu)}{\pi}\right)^3 C_FTn_l \left[ \left(
  \frac{1}{54} C_F + \frac{1}{36} C_A \right) \frac{\pi^2}{\epsilon^2} -
  \left( \frac{25}{324} C_F + \frac{37}{432} C_A \right)
  \frac{\pi^2}{\epsilon} \right] \nonumber \\
  && + \dots \,, \label{eq::Zv}
\end{eqnarray}
where the ellipses stand for non-$n_l$ and $\order{\alpha_s^4}$ terms.

Our final result for $c_v^{(3)}$ reads
\begin{eqnarray}
  c_{FFL} &=& 46.7(1) + \left( -\frac{17}{12} + \frac{61}{36} \pi^2 -
  \frac{2}{3} \pi^2 \ln 2 + \frac{1}{3} \zeta(3) \right) \Lmu +
  \frac{1}{18} \pi^2 \Lmu^2\,, \nonumber\\
  c_{FAL} &=& 39.6(1) + \left( \frac{181}{54} - \frac{67}{432} \pi^2 +
  \frac{5}{9} \pi^2 \ln 2 + \frac{13}{6} \zeta(3) \right) \Lmu +
  \left( \frac{11}{9} + \frac{1}{12} \pi^2 \right) \Lmu^2\,,
  \nonumber\\
  c_{FHL} &=& -\frac{557}{162} + \frac{26}{81} \pi^2
  + \left( -\frac{55}{27} + \frac{4}{27} \pi^2 \right) \Lmu -
  \frac{4}{9} \Lmu^2 \,, \nonumber\\
  c_{FLL} &=& -\frac{163}{162} - \frac{4}{27} \pi^2
  - \frac{11}{27} \Lmu - \frac{2}{9} \Lmu^2\,.
  \label{eq::cv3}
\end{eqnarray}
The uncertainties assigned to the numerical constants in $c_{FFL}$ and
$c_{FAL}$ are based on a conservative estimate. Note that the precision
of these quantities is more than enough for all phenomenological
applications.
Inserting the numerical values for the
colour factors we obtain for $\mu=m_Q$
\begin{eqnarray}
  c_v^{(3)} &\approx& -0.823\, n_l^2 +121.\, n_l + \mbox{non-$n_l$ terms}
  \,.
  \label{eq::cv3num}
\end{eqnarray} 
The coefficients in Eq.~(\ref{eq::cv3}) correspond to an expansion
parameter $\alpha_s(\mu)$, as given in Eq.~(\ref{eq::cvdef}). Choosing
instead $\alpha_s(m_Q)$ leads to 
\begin{eqnarray}
  \bar{c}_{FFL} &=& 46.7(1) + \frac{25}{108}\pi^2 \Lmu -
  \frac{1}{18} \pi^2 \Lmu^2\,, \nonumber\\
  \bar{c}_{FAL} &=& 39.6(1) + \frac{37}{144}\pi^2 \Lmu - \frac{1}{12}\pi^2
  \Lmu^2\,, \nonumber\\
  \bar{c}_{FHL} &=& -\frac{557}{162} + \frac{26}{81} \pi^2
  \,, \nonumber\\
  \bar{c}_{FLL} &=& -\frac{163}{162} - \frac{4}{27} \pi^2\,.
  \label{eq::cv3mq}
\end{eqnarray} 
The dependence on $L_\mu$ in Eq.~(\ref{eq::cv3mq}) is canceled against
contributions from the effective theory. They agree 
with the ones of Ref.~\cite{Kniehl:2002yv}\footnote{Note that in
  Ref.~\cite{Kniehl:2002yv} there is a typo in the coefficient of the
  $\Lmu^2$ term. The ``$-3/2$'' should read ``$+1$''.}.


\section{\label{sec::phen}Phenomenological applications}

Let us in this section estimate the numerical effect of our new terms
on the bottom and top system. In particular we consider the 
decay of the $\Upsilon(1S)$ bound state to leptons and the production
of top quark pairs close to threshold.

Next to the matching coefficient considered in the previous Sections 
a crucial ingredient for these quantities 
is the wave function at the origin. Currently the second order is
known
completely~\cite{Kuhn:1998uy,Penin:1998zh,Penin:1998kx,Penin:1998ik,Melnikov:1998ug}
and at order $\alpha_s^3$ the
quadratically~\cite{Kniehl:1999mx,Manohar:2000kr} and 
linearly~\cite{Kniehl:2002yv,Hoang:2003ns} enhanced logarithms
and the corrections proportional to
$\beta_0^3$~\cite{Penin:2005eu,Beneke:2005hg} are available. 
It is convenient to introduce the quantity
$\rho_1=|\psi_1(0)|^2/|\psi_1^C(0)|^2$ where the Coulomb wave function is
given by 
$\left|\psi^C_n(0)\right|^2=C_F^3\alpha_s^3m_q^3/(8\pi n^3)$.
Let us for completeness list the results for principle quantum number
$n=1$
\begin{eqnarray}
  \rho_1 &=& 1+\frac{\alpha_s(\mu_s)}{\pi}
  \left[\left(4-\frac{2}{3}\pi^2\right)\beta_0+\frac{3}{4}a_1\right]
  \nonumber \\
  &&{}+\left(\frac{\alpha_s(\mu_s)}{\pi}\right)^2\left\{
    \left[-C_AC_F+\left(-2+\frac{2}{3}S(S+1)\right)C_F^2\right]
  \pi^2\ln(C_F\alpha_s(\mu_s)) \right. \nonumber\\&&{}
    +\left(-\frac{5}{3}\pi^2+20\zeta(3)+\frac{1}{9}\pi^4\right)\beta_0^2
    +\left(4-\frac{2}{3}\pi^2\right)\beta_1
    +\left(\frac{5}{2}-\frac{2}{3}\pi^2\right)\beta_0a_1
    \nonumber \\&&{}
    +\frac{3}{16}a_1^2+\frac{3}{16}a_2
    +\left.\frac{9}{4}\pi^2C_AC_F
    +\left(\frac{33}{8}-\frac{13}{9}S(S+1)\right)\pi^2C_F^2\right\}
    \nonumber\\&&{}
    +\left(\frac{\alpha_s(\mu_s)}{\pi}\right)^3
    \bigg\{ \pi^2 {\cal C}_2 \ln^2\left(C_F\alpha_s(\mu_s)\right)
    +\pi^2 {\cal C}_1\ln\left(C_F\alpha_s(\mu_s)\right)
    +{\cal C}_0^{\beta_0^3} 
    +\ldots
    \bigg\}
    \,,
    \label{eq::rho1}
\end{eqnarray}
with 
\begin{eqnarray}
  {\cal C}_2 &=&
  \left(-2C_AC_F+
  \left(-4+\frac{4}{3}S(S+1)\right)C_F^2\right)\beta_0
  -\frac{2}{3}C_A^2C_F
  \nonumber\\&&{}\mbox{}
  +\left(-\frac{41}{12}+\frac{7}{12}S(S+1)\right)C_AC_F^2
  -\frac{3}{2}C_F^3
  \,,
\end{eqnarray}
\begin{eqnarray}
  {\cal C}_1&=&\left[\left(-3+\frac{2}{3}\pi^2\right)C_AC_F+
    \left(\frac{4}{3}\pi^2-\left(\frac{10}{9}+\frac{4}{9}\pi^2
    \right) S(S+1)\right) C_F^2\right]\beta_0
  \nonumber\\&&{}
  +\left[-\frac{3}{4}C_AC_F+\left(-\frac{9}{4}
    +\frac{2}{3}S(S+1)\right) C_F^2\right]a_1
  +\frac{1}{4}C_A^3 + \left(\frac{59}{36}-4\ln2\right)C_A^2C_F
  \nonumber \\&&{}
  +\left(\frac{143}{36} - {4}\ln 2-\frac{19}{108} S(S+1)\right)C_AC_F^2
  +\left(-\frac{35}{18}+8\ln2-\frac{1}{3} S(S+1)\right)C_F^3
\nonumber\\&&{}
  +\left(-\frac{32}{15}+2\ln2+\left(1-\ln2\right)S(S+1)\right) C_F^2T
  +\frac{49}{36}C_AC_FTn_l
\nonumber\\&&{}
  +\left(\frac{8}{9}-\frac{10}{27}S(S+1)\right)C_F^2Tn_l\,
  \label{cln}
\end{eqnarray}
and
\begin{eqnarray}
  {\cal C}_0^{\beta_0^3} &=&\beta_0^3
  \Bigg[
    -{20}+\frac{22}{3}\pi^2 + 112\zeta(3) - \frac{7}{5}\pi^4 -
  {12\pi^2\zeta(3)} - 40\zeta(5) - 16\zeta(3)^2
  \nonumber\\&&\mbox{}
  + \frac{4}{105}\pi^6
  \Bigg]
  \,,
\end{eqnarray}
where $\mu_s=C_F m_q\alpha_s(\mu_s)$ is the soft scale. 
It is straightforward to obtain the result for general $\alpha_s(\mu)$
using standard renormalization group analyses.
In Eqs.~(\ref{eq::rho1})--(\ref{cln}) $S$ is the spin quantum number which is equal
to one in our applications. The ellipses in Eq.~(\ref{eq::rho1})
represent yet unknown corrections like, e.g., the pure ultrasoft
contributions. 

For completeness we provide the
one- and two-loop coefficients of the $\beta$ function and the static
potential which are given by
\begin{eqnarray}
  \beta_0 &=& \frac{1}{4} \left( \frac{11}{3} C_A - \frac{4}{3} Tn_l
  \right)
  \,,\nonumber\\
  \beta_1 &=& \frac{1}{16} \left( \frac{34}{3} C_A^2 - \frac{20}{3}
  C_ATn_l - 4 C_FTn_l \right)
  \,,\nonumber\\
  a_1 &=& \frac{31}{9} C_A - \frac{20}{9} Tn_l
  \,,\nonumber\\
  a_2 &=& \left( \frac{4343}{162} + 4 \pi^2 + \frac{22}{3} \zeta(3) -
  \frac{1}{4} \pi^4 \right) C_A^2 - \left( \frac{1798}{81} +
  \frac{56}{3} \zeta(3) \right) C_ATn_l \nonumber\\
  && - \left( \frac{55}{3} - 16 \zeta(3) \right) C_FTn_l + \left(
  \frac{20}{9} Tn_l \right)^2
  \,.
\end{eqnarray}

\subsection{Bottom system}

The leptonic decay of the $\Upsilon(1S)$ state can be cast in the
form~\cite{Kuhn:1998uy,Penin:1998zh,Penin:1998kx,Penin:1998ik,Melnikov:1998ug} 
\begin{eqnarray}
  \Gamma(\Upsilon(1S)\to l^+ l^-) &=& \Gamma^{\rm LO} \rho_1
  \left[c_v^2(m_b)+\frac{C_F^2\alpha_s^2(\mu_s)}{12}c_v(m_b)\left(d_v(m_b)+3\right)
    \right]
  + \ldots
  \,,
  \label{eq::gamll}
\end{eqnarray}
with $\Gamma^{\rm LO}=4\pi
N_cQ_b^2\alpha^2|\psi_1^C(0)|^2/\left(3m_b^2\right)$, $Q_b=-1/3$, $N_c =
3$, $\alpha$ is Sommerfeld's fine-structure constant
and nonperturbative contributions to Eq.~(\ref{eq::gamll}) are ignored.

Inserting the perturbative expansion for $\rho_1$ and $c_v$ we obtain
\begin{eqnarray}
  \Gamma_1 
  &\approx& \Gamma_1^{\rm LO}
  \left(1
  -1.70\,\alpha_s(m_b)
  -7.98\,\alpha_s^2(m_b)
  + 30.0\,\alpha_s^3(m_b)|_{n_l}
  +\ldots\right)
  \nonumber\\
  &&{}\times\Big[ 1
  -0.30\,\alpha_s(\mu_s)
  + \alpha_s^2(\mu_s) \left( 17.2 -5.19 \ln\alpha_s(\mu_s) \right)
  \nonumber\\&&{}\left.
  + \alpha_s^3(\mu_s) \left(
  -14.4 \ln^2\alpha_s(\mu_s)
  +0.17 \ln\alpha_s(\mu_s)
  -34.9 |_{\beta_0^3} \right)
  +\ldots\right]
  \,,
  \label{eq::gamser}
\end{eqnarray}
where in the matching coefficient $\mu=m_b$ has been chosen and 
the corresponding strong coupling is defined with five and
$\alpha_s(\mu_s)$ with four active flavours.

Starting from $\alpha_s(M_Z)=0.118$ we have used the program {\tt
  RunDec}~\cite{Chetyrkin:2000yt} to obtain
$\alpha_s(m_b)=0.2096$ and $\alpha_s(\mu_s)=0.2967$
with $m_b=5.3$~GeV and $\mu_s=2.0967$~GeV which leads to 
\begin{eqnarray}
  \Gamma_1 &\approx& \Gamma_1^{\rm LO}(1 -0.446_{\rm NLO}
  +1.75_{\rm NNLO}-1.20_{\rm N^3LO^\prime}+\ldots)
  \,,
  \label{eq::gamnum}
\end{eqnarray}
where Eq.~(\ref{eq::gamser}) is expanded and terms of order
$\alpha_s^4$ are dropped consistently. The prime reminds that the
third-order corrections are not complete. Apart form the new
contribution to $c_v$ and the known third-order corrections to $\rho_1$
we have also included all interference terms which are proportional
to powers of $n_l$.
Note that the new corrections are responsible for the
reduction of N$^3$LO$^\prime$ terms from $-1.67$ to $-1.20$ which
amounts to about 47\% of the Born cross section.
However, in total the fermionic corrections tend to further reduce the
strong increase of the perturbative coefficients leading to an overall
correction factor of approximately 10\%.

We have performed the numerical analysis also for $m_b=4.8$~GeV. There
are changes in the individual contributions to $\Gamma_1$
(c.f. Eq.~(\ref{eq::gamnum})) of the order of a few 
per cent, however, the final correction factor remains the same.

\subsection{Top system}

In the top-quark case, the nonperturbative effects are negligible.
However, the effect of the top-quark total decay width $\Gamma_t$ has to be
properly taken into account \cite{Fadin:1987wz}, as it is relatively large and
smears out the Coulomb-like resonances below threshold.
The NNLO analysis of the cross
section \cite{Hoang:2000yr} shows that only the ground-state pole gives rise to a
prominent resonance.

A crucial quantity in connection to the threshold production of top
quark pairs is the peak of the normalized cross section  
$R=\sigma(e^+e^-\to t\bar t)/\sigma(e^+e^-\to\mu^+\mu^-)$.
It is dominated by the contribution from the would-be toponium
ground-state, which can be cast in the from
\begin{eqnarray}
  R_1(e^+e^-\to t\bar{t}) &=& R_1^{\rm LO}\rho_1
  \left[c_v^2(m_t)+\frac{C_F^2\alpha_s^2(\mu_s)}{12}c_v(m_t)\left(d_v(m_t)+3\right)
    \right]
  + \ldots
  \,,
  \label{eq::ree}
\end{eqnarray}
with the leading order term
$R_1^{\rm LO}=6\pi N_cQ_t^2|\psi_1^C(0)|^2/\left(m_t^2
\Gamma_t\right)$.
The contributions from the higher Coulomb-like poles and the continuum are not
included in Eq.~(\ref{eq::ree}).

The analog equations to Eqs.~(\ref{eq::gamser}) and~(\ref{eq::gamnum})
read
\begin{eqnarray}
  R_1   &\approx& R_1^{\rm LO}
  \left(1
  -1.70\,\alpha_s(m_t)
  -7.89\,\alpha_s^2(m_t)
  + 37.2\,\alpha_s^3(m_t)|_{\rm n_l}
  +\ldots\right)
  \nonumber\\
  &&{}\times\Big[1
  -0.43\,\alpha_s(\mu_s)
  + \alpha_s^2(\mu_s) \left( 16.1 - 5.19 \ln\alpha_s(\mu_s) \right)
  \nonumber\\&&{}\left.
  + \alpha_s^3(\mu_s) \left(
  -13.8 \ln^2\alpha_s(\mu_s)
  +2.06 \ln\alpha_s(\mu_s)
  -27.2 |_{\beta_0^3} \right)
  +\ldots\right]
  \,,
  \label{eq::rser}
\end{eqnarray}
where $\alpha_s(m_t)$ and $\alpha_s(\mu_s)$ are defined with six and five
active flavours, respectively. Using again $\alpha_s(M_Z)=0.118$ 
one gets $\alpha_s(m_t)=0.1075$
and $\alpha_s(\mu_s)=0.1398$ with
$m_t=175$~GeV and $\mu_s=32.625$~GeV and finally
\begin{eqnarray}
  R_1 &\approx& R_1^{\rm LO}(1 -0.243_{\rm NLO}
  +0.435_{\rm NNLO}-0.195_{\rm N^3LO^\prime}+\ldots)
  \,.
  \label{eq::rnum}
\end{eqnarray}
The fermionic corrections to $c_v$ are responsible for a
reduction of the third-order coefficient from $-0.268$ to
$-0.195$ and thus amount to moderate 7\% of the leading order term.
Similarly as for the bottom quark case also for the top quark the
perturbative series is alternating and the third-order coefficient 
tends to stabilize the expansion.
It is interesting to note that after the inclusion of our new
terms the total corrections amount to less than 1\%.


\section{\label{sec::concl}Conclusions and outlook}

This paper deals with the question of establishing a relation for the
vector current between full QCD and NRQCD. The corresponding matching
coefficient, $c_v$, constitutes a building block in all threshold phenomena
involving a vector coupling.
The main result of this paper is the fermionic 
three-loop contribution to $c_v$ which contains a closed light fermion
loop. The numerical effect of the new terms is relatively big and 
amounts to about 47\%
for the bottom and to about 7\% for the top quark system.
However, their inclusion leads to a reduction of the
overall corrections. E.g., in the case of the top quark the corrections
to the normalization at the peak of the production cross section
amount to less than 1\% after including all currently known perturbative terms. 

Several steps still have to be taken in order to arrive at the
complete analysis of the top-anti-top quark system at threshold. Next
to the non-fermionic pieces to $c_v$ the major building blocks which
are still missing are the ultra-soft corrections and the three-loop
static potential.


\vspace*{1em}

\noindent
{\large\bf Acknowledgements}\\
We would like to thank A.A.~Penin and V.A.~Smirnov for useful comments
and discussions.
J.H.P. would like to thank S.~Bekavac for discussions about
Mellin-Barnes integrals and  M.~Kalmykov for useful advice on two-loop
sunset integrals. We thank M.~Tentyukov for providing the interface to
{\tt Fermat}.
This work was supported by the ``Impuls- und Vernetzungsfonds'' of the
Helmholtz Association, contract number VH-NG-008 and 
by the DFG through SFB/TR~9. The Feynman diagrams were
drawn with {\tt JaxoDraw} \cite{Binosi:2003yf}


\begin{appendix}

\section{Results for the master integrals\label{masters}}
\begin{figure}
  \leavevmode
  \epsfxsize=\textwidth
 \epsffile{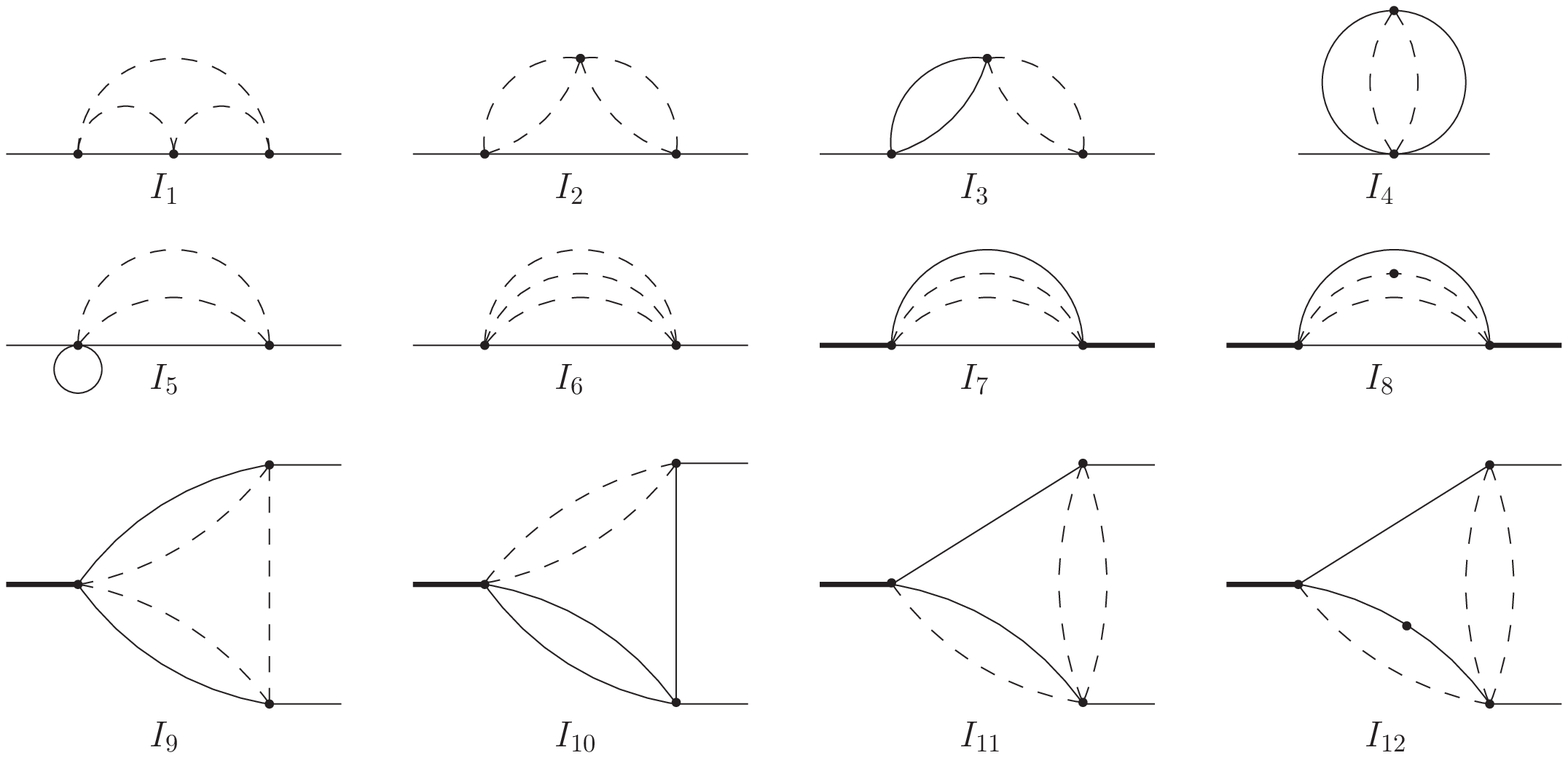}
\caption{\label{fig::masters} Three-loop master integrals. Bold lines
  denote massive lines with mass $2m_Q$, thin lines denote massive lines
  with mass $m_Q$ and dashed lines denote massless lines. All external
  lines are on-shell. A dot on a line denotes a squared propagator.}
\end{figure}

The master integrals for the three-loop functions of
Eq.~(\ref{eq::integrals}) are shown in Fig.~\ref{fig::masters}. For the
self-energy integrals $I_1$--$I_6$ the external momentum squared equals
the square of the heavy quark mass, $q^2 = m_Q^2$. For the integrals
$I_7$ and $I_8$ we have $q^2 = 4 m_Q^2$.
The momenta $q_1$ and $q_2$ flowing out of the vertex diagrams on the
right fulfill $q_1^2 = q_2^2 = m_Q^2$ with $(q_1 + q_2)^2 = 4 m_Q^2$.
In Minkowski space the results for the master integrals read
\begin{eqnarray}
  I_1 &=& J_+^{(3)}(0,1,0,1,1,0,1,0,1) \nonumber \\
  &=& m_Q^2 \left( \frac{\mu^2}{m_Q^2}  e^{-\gamma_E}\right)^{3 \epsilon}
  \left\{ \frac{1}{3\epsilon^3} +
  \frac{5}{3\epsilon^2} + \left(4 + \frac{3}{4}\pi^2\right)
  \frac{1}{\epsilon} - \frac{10}{3} + \frac{11}{4} \pi^2 + \frac{25}{3}
  \zeta(3) \right. \nonumber \\
  && + \left( -\frac{302}{3} + 2\pi^2 + \frac{89}{3}\zeta(3) +
  \frac{3059}{1440} \pi^4 \right)\epsilon + \left( -734 -
  \frac{69}{2}\pi^2 + 16\zeta(3) + \frac{9751}{1440} \pi^4
  \right. \nonumber \\
  && \left.\left. + \frac{107}{4}\pi^2\zeta(3) +
  \frac{2309}{5}\zeta(5) \right) \epsilon^2 + \order{\epsilon^3}
  \right\} \,,
\\
  I_2 &=& J_+^{(3)}(0,1,1,1,0,1,1,0,0) \nonumber \\
  &=& -m_Q^2 \left( \frac{\mu^2}{m_Q^2} \right)^{3 \epsilon}
  \frac{\Gamma^4(1-\epsilon) \Gamma^2(\epsilon)}{\Gamma^2(2-2\epsilon)}
  \frac{\Gamma(3\epsilon-1) \Gamma(3-6\epsilon)}{\Gamma(3-4\epsilon)} \,,
\end{eqnarray}
\begin{eqnarray}
  I_3 &=& L_+^{(3,n_l)}(0,0,1,1,1,1,0,1,0) \nonumber \\
  &=& m_Q^2 \left( \frac{\mu^2}{m_Q^2} e^{-\gamma_E}\right)^{3 \epsilon}
  \left\{ \frac{2}{3\epsilon^3} +
  \frac{10}{3\epsilon^2} + \left( \frac{26}{3} + \frac{1}{2}\pi^2 \right)
  \frac{1}{\epsilon} + 2 + \frac{9}{2}\pi^2 + \frac{14}{3}\zeta(3)
  \right. \nonumber \\
  && + \left( -\frac{398}{3} + \frac{53}{2}\pi^2 - 16\pi^2 \ln 2 +
  \frac{238}{3}\zeta(3) + \frac{287}{720}\pi^4 \right) \epsilon
  + \left( -1038 + \frac{259}{2}\pi^2 \right. \nonumber \\
  && - 160\pi^2 \ln 2 + \frac{128}{3} \pi^2 \ln^2 2 +
  \frac{1862}{3}\zeta(3) + \frac{323}{720}\pi^4 +
  \frac{7}{2}\pi^2\zeta(3) + \frac{478}{5}\zeta(5) \nonumber \\
  && \left. + \frac{64}{3}\ln^4 2 + 512 a_4 \right) \epsilon^2 +
  \order{\epsilon^3} \bigg\} \,,
\\
  I_4 &=& J_+^{(3)}(0,0,0,1,1,0,1,0,1) \nonumber \\
  &=& m_Q^4 \left( \frac{\mu^2}{m_Q^2} \right)^{3 \epsilon}
  \frac{\Gamma^2(1-\epsilon) \Gamma(\epsilon) \Gamma^2(2\epsilon-1)
  \Gamma(3\epsilon-2)}{\Gamma(4\epsilon-2) \Gamma(2-\epsilon)} \,,
\\
  I_5 &=& J_+^{(3)}(0,1,0,1,0,0,1,0,1) \nonumber \\
  &=& m_Q^4 \left( \frac{\mu^2}{m_Q^2} \right)^{3 \epsilon}
  \Gamma(\epsilon-1) \frac{\Gamma^2(1-\epsilon)
  \Gamma(\epsilon)}{\Gamma(2-2\epsilon)} \frac{\Gamma(2\epsilon-1)
  \Gamma(3-4\epsilon)}{\Gamma(3-3\epsilon)} \,,
\\
  I_6 &=& J_+^{(3)}(0,0,1,1,1,0,1,0,0) \nonumber \\
  &=& m_Q^4 \left( \frac{\mu^2}{m_Q^2} \right)^{3 \epsilon}
  \frac{\Gamma^3(1-\epsilon) \Gamma(2\epsilon-1) \Gamma(3\epsilon-2)
  \Gamma(5-6\epsilon)}{\Gamma(3-3\epsilon) \Gamma(4-4\epsilon)} \,,
\\
  I_7 &=& J_-^{(3)}(0,0,0,1,1,0,1,0,1) \nonumber \\
  &=& m_Q^4 \left( \frac{\mu^2}{m_Q^2} e^{-\gamma_E} \right)^{3 \epsilon}
  \left\{ \frac{1}{3\epsilon^3} +
  \frac{1}{2\epsilon^2} + \left(-\frac{17}{36} + \frac{1}{12} \pi^2
  \right)\frac{1}{\epsilon} - 6.6827387(1) \right. \nonumber \\
  && - 56.300353(1)\epsilon - 209.48231(1)\epsilon^2 + \order{\epsilon^3}
  \bigg\} \,,
\\
  I_8 &=& J_-^{(3)}(0,0,0,2,1,0,1,0,1) \nonumber\\
  &=& -m_Q^2 \left( \frac{\mu^2}{m_Q^2} e^{-\gamma_E} \right)^{3 \epsilon}
  \left\{ \frac{1}{3\epsilon^3} +
  \frac{2}{3\epsilon^2} + \left( -\frac{8}{3} + \frac{5}{6}\pi^2 \right)
  \frac{1}{\epsilon} + 10.797602(1)
  \right. \nonumber\\
  && + 62.250613(1)\epsilon + \order{\epsilon^2} \bigg\} \,,
\\
  I_9 &=& J_-^{(3)}(0,1,0,1,1,0,1,0,1) \nonumber \\
  &=& m_Q^2 \left( \frac{\mu^2}{m_Q^2} e^{-\gamma_E} \right)^{3 \epsilon}
  \left\{ \frac{1}{3\epsilon^3} +
  \frac{5}{3\epsilon^2} + \left( \frac{10}{3} + \frac{3}{4}\pi^2 \right)
  \frac{1}{\epsilon} + 33.8328(4) + 152.870(4)\epsilon \right. \nonumber \\
  &&+ \order{\epsilon^2} \bigg\} \,,
\end{eqnarray}
\begin{eqnarray}
  I_{10} &=& L_+^{(3,n_l)}(0,0,1,1,1,0,1,1,0) \nonumber\\
  &=& m_Q^2 \left( \frac{\mu^2}{m_Q^2} e^{-\gamma_E} \right)^{3 \epsilon}
  \left\{ \frac{2}{3\epsilon^3} +
  \frac{10}{3\epsilon^2} + \left(8 + \frac{1}{2} \pi^2 \right)
  \frac{1}{\epsilon} + 52.5698(4) + 145.087(4)\epsilon \right. \nonumber \\
  && + 562.250(14)\epsilon^2 + \order{\epsilon^3}
  \bigg\}  \,,
\\
  I_{11} &=& J_-^{(3)}(0,1,0,1,0,1,1,0,1) \nonumber \\
  &=& m_Q^2 \left( \frac{\mu^2}{m_Q^2} e^{-\gamma_E} \right)^{3 \epsilon}
  \left\{ \frac{1}{2\epsilon^3} +
  \frac{11}{6\epsilon^2} + \left( -\frac{1}{6} + \frac{29}{24}\pi^2
  \right) \frac{1}{\epsilon} + 34.791(4) + 154.08(2)\epsilon
  \right. \nonumber \\
  &&+ \order{\epsilon^2} \bigg\} \,,
\\
  I_{12} &=& J_-^{(3)}(0,1,0,1,0,1,1,0,2) \nonumber \\
  &=& \left( \frac{\mu^2}{m_Q^2} e^{-\gamma_E} \right)^{3\epsilon}
  \left\{ \frac{1}{6\epsilon^3} + \frac{1}{2\epsilon^2} +
  \left( \frac{1}{6} + \frac{7}{24}\pi^2 \right) \frac{1}{\epsilon} +
  7.024(4) + 32.04(2)\epsilon \right. \nonumber \\
  && + \order{\epsilon^2} \bigg\} \,,
\end{eqnarray}
where $\zeta(5) = 1.036927\ldots$ and $a_4 = {\rm Li}_4(1/2) =
0.517479\ldots$.

$I_1$ was calculated in Ref.~\cite{Melnikov:2000zc}. We have repeated
the calculation described in the reference with the program
\texttt{XSummer}~\cite{Moch:2005uc} and find complete agreement.  The
results for $I_3$--$I_6$ can be found in Ref.~\cite{Laporta:1996mq}. We
have checked the result for $I_3$ numerically with the help of a one-fold
Mellin-Barnes~\cite{Smirnov:1999gc,Tausk:1999vh} representation and find
complete agreement. Note, that $I_2, I_4, I_5$ and $I_6$ are quite
simple to evaluate and are available for general $\epsilon$.
Nevertheless the result for $I_6$ as given in
Ref.~\cite{Melnikov:2000zc} is not correct whereas the correct result
can be found in Ref.~\cite{Laporta:1996mq}. The pole part of $I_7$
agrees with the result of Ref.~\cite{Groote:2004qq}.

The remaining master integrals, $I_7$--$I_{12}$, have been evaluated
with the help of the Mellin-Barnes method where the 
evaluation of the integrals has been performed with the program
\texttt{MB}~\cite{Czakon:2005rk}.
$I_7$ and $I_8$ can be expressed in terms of a one-fold Mellin-Barnes
representation and are thus known with a quite high precision. On the
other hand, $I_{10}$ is represented by a two-fold and 
$I_9$, $I_{11}$ and $I_{12}$ even by a three-fold integration which
results in less accurate results. The quoted uncertainties in the above
equations correspond to twice the Vegas error given by {\tt MB} for the
multi-dimensional integrals and to a conservative estimate in case of
the one-dimensional integrals $I_7$ and $I_8$. Note that the latter
errors are negligible.

\begin{figure}
 \centerline{\epsfig{file=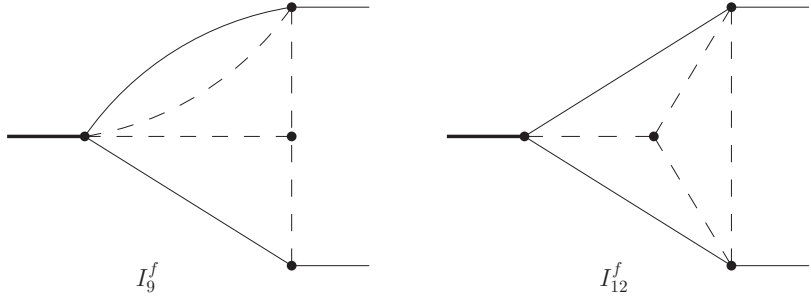,scale=0.65}}
\caption{\label{fig::epfin} $\epsilon$-finite master integrals. The same
coding as in Fig.~\ref{fig::masters} is adopted.}
\end{figure}

As can be seen in the above results some of the master integrals are
needed to higher order in $\epsilon$ which on one hand makes the
calculation very tedious and on the other hand leads to less accurate
results. For this reason we decided to change basis and switch --- at
least for some of the integrals --- to the so-called
$\epsilon$-finite master integrals~\cite{Chetyrkin:2006dh} which have
the advantage that the 
coefficient function is finite and thus the integral itself is only
needed to order $\epsilon^0$.

Since $I_1$--$I_6$ are known analytically, a replacement is, of
course, not necessary. Furthermore, for $I_7$ and $I_8$ the numerical
precision is 
sufficient for our calculation. 
As far as the remaining four integrals are concerned, we
found it convenient to replace $I_9$ and $I_{12}$ by the integrals
shown in Fig.~\ref{fig::epfin}. 
Their numerical evaluation with the help of \texttt{MB} is
straightforward leading to the results
\begin{eqnarray}
  I_9^f &=& J_-^{(3)}(0,1,1,1,1,0,1,0,1) \nonumber \\
  &=& \left( \frac{\mu^2}{m_Q^2}  e^{-\gamma_E}\right)^{3\epsilon}
  \left\{ \frac{1}{6\epsilon^3} +
  \frac{3}{2\epsilon^2} + \left( \frac{55}{6} + \frac{3}{8} \pi^2
  \right) \frac{1}{\epsilon} + 64.678(8) + \order{\epsilon} \right\} \,,
\\
  I_{12}^f &=& J_-^{(3)}(0,1,0,1,1,1,1,0,1) =  \left(
  \frac{\mu^2}{m_Q^2}  e^{-\gamma_E}\right)^{3\epsilon}
  \left\{ \frac{2.4041(4)}{\epsilon} + 8.1(2) + \order{\epsilon} \right\} \,.
\end{eqnarray}
Switching from $I_9$ and $I_{12}$ to $I_9^f$ and $I_{12}^f$ reduces
the number of coefficients which are only known numerically from
17 to 14. In particular the $1/\epsilon$-poles of $I_9$ and
$I_{12}$ have been determined in analytical form from the
$\epsilon$-finite integrals via the corresponding IBP relations.

A further reduction is achieved after exploiting the locality and
incorporating the knowledge of the linear $\ln(\mu^2/m_Q^2)$ term in
$c_v$~\cite{Kniehl:2002yv} which we have checked numerically in a first
step. In this way the analytical result for the $1/\epsilon$-pole
of $I_{11}$ has been determined. In total ten (eleven) numerical
coefficients contribute to our final result given in
Eqs.~(\ref{eq::cv3}) and~(\ref{eq::cv3num}) using the $\epsilon$-finite
(``normal'') basis.

In the end it turned out, that the numerical precision is better using
the ``normal'' basis. The reason is, that the uncertainty of the
$\epsilon^0$-coefficient of $I_{12}^f$ is by far the largest. Still, it
was useful to consider the $\epsilon$-finite basis since it enabled us
to determine analytical results for two integral coefficients of the
``normal'' basis.

\end{appendix}



\end{document}